\documentstyle[preprint,pre,aps,eqsecnum,epsfig]{revtex}

\begin{document}
\newcommand{\be}{\begin{equation}}
\newcommand{\ee}{\end{equation}}
\newcommand{\ba}{\begin{eqnarray}}
\newcommand{\ea}{\end{eqnarray}}
\preprint{PISA IFUP-TH 17, March 1994}
\draft
\title{
Bosonization and the lattice:\\
the $d=2$ Gross-Neveu model}
\author{Matteo Beccaria$^{(1,2)}$}
\address{
(1) Dipartimento di Fisica, Universit\'a di Pisa\\
Piazza Torricelli 2, I-56100 Pisa, Italy\\
(2) I.N.F.N., sez. di Pisa\\
Via Livornese 582/a, I-56010 S. Piero a Grado (Pisa) Italy
}
\date{\today}
\maketitle
\begin{abstract}
We consider the lattice formulation of bosonized quantum field theories. As a
non
trivial example, we study the two dimensional Gross-Neveu model.
Analytical investigations and direct numerical
simulation strongly suggest that the lattice model reproduces the continuum
physics.
Anticommuting fields are not required and there are not related doubling
problems.
\end{abstract}
\vskip 1truecm
\pacs{PACS numbers: 11.15.Ha, 05.50.+q, 11.30.Na, 11.30.Rd}
\section{Introduction}
\label{sec:intro}
In a recent work~\cite{Beccaria:GNB}, we proposed a lattice action for the
bosonized
$N$-flavors two dimensional Gross-Neveu model. In
principle, the possibility of avoiding fermionic fields is very interesting
from the point of view of
numerical simulations. Indeed, current Monte Carlo simulations of dynamical
fermions are hampered
by problems like the inversion of the fermionic propagator or the unavoidable
breaking of
chiral symmetry predicted by the Nielsen-Ninomiya theorem~\cite{Nielsen}.
In the past, the powerful technique of bosonization remained somewhat limited
to $1+1$
dimensions and to specific models.
{}From a modern point of view, bosonization is better described as a a
particular case of
duality and has thus a wider
range of applicability~\cite{Burgess}.

Can this equivalence be carried over to the lattice ? Actually, one can
bosonize the algebra of
fermionic lattice fields but this construction is usually based on the
Jordan-Wigner
transformation (see \cite{Hooft} for a very interesting example) which is non
local and leads to
complicate simulation algorithms.
In~\cite{Beccaria:GNB}, we decided to bosonize in the continuum and to put the
resulting non
polynomial model on the lattice. The simulation algorithms are the usual ones
since the computer
does not care of non-polinomiality. The question is therefore if a critical
point can be found and
if it does describe the continuum physics of the original fermionic model.

The analysis in~\cite{Beccaria:GNB} was preliminary and the numerical results
concerned mainly
the $1/N$ expansion of the lattice model. No informations were obtained
regarding doubling
and scaling. Moreover the bound states sector was not studied. In this work we
report on additional
numerical results in order to check the recovery of continuum asymptotic
scaling including the
measure of the lightest bosonic state.
The plan of the paper is the following: in Sec.~\ref{sec:boso} we review the
bosonization formalism and
explain how to apply it to the Gross Neveu model. In Sec.~\ref{sec:latt} we
recall the lattice action
and discuss its relation with the usual formulation. Sec.~\ref{sec:exp} is
devoted to the $1/N$
expansion. In Sec.~\ref{sec:RG} we collect results on Renormalization Group
scaling and the
analytical predictions on the mass spectrum. Finally, Sec.~\ref{sec:num}
describes the simulation
and the results. The appendix contains some technical remarks
concerning the Symanzik improvement of the model.
\section{The bosonization technique}
\label{sec:boso}
It is a very old result~\cite{Coleman1,Mandelstam} that in two space-time
dimensions a free massless fermion is equivalent to a
free massless boson. The equivalence can be proved case by case in various
interacting models too.
Indeed, one can write down a mapping, the Mandelstam
representation~\cite{Mandelstam}, which allows
to solve the fermionic
equations of motion in terms of bosonic operators. One can renormalize in order
to obtain the
same Green  functions for bilinear local fermionic operators and suitable local
bosonic ones.
Recently, a very interesting connection has been shown between
bosonization and duality which allows a more constructive
approach~\cite{Burgess}.
The original fermionic theory is embedded in a larger gauge model, the field
strength is made to
vanish with the aid of a suitable Lagrange multiplier and, finally, by
integrating out the original
and the gauge fields, the effective bosonized action for the multiplier is
obtained. As far as non
abelian symmetry of the Fermi multiplets is allowed to be non linearly realized
in the bosonic
representation, the Mandelstam construction is
\ba
\psi_L &=&  \sqrt{\frac{c\mu}{2\pi}}
: {\rm exp} \left\{-i\sqrt \pi\left(\int^x_{-\infty}d\xi\ \pi(\xi)+\varphi
(x)\right)\right\} :, \\
\psi_R &=&  \sqrt{\frac{c\mu}{2\pi}}
: {\rm exp}\left\{-i\sqrt\pi\left(\int^x_{-\infty} d\xi\ \pi(\xi)-\varphi(x)
\right)\right\} : \nonumber
\ea
where $\psi_{L,R}$ are chiral Weyl fermions, $\varphi(x)$ is a scalar massless
field,
$\pi(x) = \partial_0\varphi(x)$ and $c = \frac{1}{2}e^\gamma$.
Normal ordering is with respect to the mass $\mu$~\cite{Coleman2}.
The rules of bosonization are therefore
\be
\label{dict}
\bar{\psi}{\partial \!\!\! /}\psi\to \frac{1}{2} (\partial_\mu\varphi)^2\qquad
\bar{\psi}\psi\to \frac{c\mu}{\pi} :\cos\beta\varphi : \qquad \beta^2 = 4\pi.
\ee
A straigthforward application of this formalism to the Gross-Neveu model is
possible~\cite{WittenShankar}, but not suitable for a
successive expansion in powers of $1/N$. The continuum action is
\begin{equation}
\label{without}
S = \int d^2x \left({\bar\psi}^{(\alpha)}
{\partial \!\!\! /}
\psi^{(\alpha)}-\frac{1}{2} g^2
\left({\bar\psi}^{(\alpha)}\psi^{(\alpha)}\right)^2 \right) \qquad \alpha =
1\cdots N,
\end{equation}
where $\psi^{(\alpha)}$ is a multiplet of two-dimensional spinors.
Our idea is to apply Eqs.(\ref{dict}) to the action
\begin{equation}
\label{with}
S = \int d^2x \left({\bar\psi}^{(\alpha)}
{\partial \!\!\! /}
\psi^{(\alpha)}+\sigma
{\bar\psi}^{(\alpha)}\psi^{(\alpha)}+\frac{1}{2g^2}\sigma^2
\right)
\end{equation}
which differs from Eq.(\ref{without}) only for the introduction of a Lagrange
multiplier
which eliminates the quartic interaction.
It is not obvious that one could apply to Eq.(\ref{with})
the bosonization dictionary as if the $\sigma\bar{\psi}\psi$ were a mass term.
Furthermore, the
equivalence of the bosonized theory and the original one is proved only at a
perturbative level.
The purpose of the present investigation is just to understand which critical
point is reached by
the Renormalization Group flow of the lattice model we are going to study.
\section{The lattice action and its properties}
\label{sec:latt}
For later study of the $1/N$ expansion, it is convenient to write down the
usual
formulation and the bosonized one with a similar notation. We shall see that
the
formulae in~\cite{Curci} can be immediately adapted in both frameworks. The
usual model which
is studied on the lattice as a discrete version of the Gross-Neveu model is
defined by
\ba
S &=& \sum_n\frac{N}{2\lambda}\sigma_n^2 + S_{\rm kin}(\psi) + S_{\rm
int}(\sigma, \psi), \nonumber \\
S_{\rm kin}(\psi) &=& \sum_{n,m,\alpha} \bar{\psi}^\alpha_n M_{nm}
\psi_m^\alpha, \\
S_{\rm int}(\sigma, \psi) &=& \sum_{n,\alpha}\ \bar{\psi}^\alpha_n\
\psi_n^\alpha\ \sigma_n,
\nonumber \ea
where $\psi^{(\alpha)}$ is a set of Dirac fermions and $M$ is the inverse
lattice propagator
supplemented with suitable terms needed to avoid doubling, for instance, we can
think to the
Wilson proposal~\cite{Curci}. The corresponding
bosonic model is
\ba
S &=& \sum_n\frac{N}{2\lambda}\sigma_n^2 + S_{\rm kin}(\varphi) + S_{\rm
int}(\sigma, \varphi),
\nonumber \\
S_{\rm kin}(\varphi) &=& \frac{1}{2}\ \Delta_\mu^+\varphi_n^\alpha\
\Delta_\mu^+\varphi_n^\alpha,
\qquad \Delta_\mu^+\varphi_n^\alpha =
\varphi_{n+\hat{\mu}}^\alpha-\varphi_n^\alpha, \\
S_{\rm int}(\sigma, \varphi) &=& -\omega\sum_{n,\alpha}
\sigma_n\cos\beta\varphi_n^\alpha,
\nonumber\ea
where $\varphi^{(\alpha)}$ is a set of $N$ massless bosonic fields.
The constant $\omega$ is related to the normal product appearing in
Eqs.(\ref{dict}),
and is determined by the value of the tadpole
\be
\langle \cos\beta\varphi_n \rangle  = e^{-\frac{1}{2}\beta^2 \Delta_{n,n}}
\ee
where $\Delta = M^{-1}$ is the lattice propagator. We take
\be
\omega = \lim_{\mu\to 0}\frac{c\mu}{\pi} \exp\frac{\beta^2}{2}\int\frac{d^2
p}{(2\pi)^2}
\frac{1}{\hat{p}^2 + \mu^2} = \frac{c\ 2^{5/2}}{\pi}
\ee
in order to reproduce the continuum expression
\be
\mu\ :\cos\beta\varphi: \qquad \mbox{normal order at mass}\ \mu
\ee
in the $a\to 0$ limit, $a$ being the lattice spacing.
The model has a discrete chiral symmetry which acts on the fermionic fields as
\be
\psi\to \gamma_5\psi\qquad\bar{\psi}\to -\bar{\psi}\gamma_5\qquad
\sigma\to -\sigma .
\ee
and on the bosons
\be
\sigma\to-\sigma \qquad\qquad \varphi\to\varphi+\frac{\pi}{\beta} .
\ee
This symmetry is dynamically broken: loop corrections change the shape of the
effective potential
for the chiral condensate ($\bar{\psi}\psi$ related to $\sigma$ by the
Schwinger-Dyson equations)
which develops a non zero vacuum expectation value. On the lattice there is
also an explicit breaking
with fermions, because of the Wilson term needed to avoid doublers. In the
bosonic formulation,
the symmetry is not broken by the action and the dynamical breaking chooses the
two $Z_2$ vacua
with equal chances. Periodic boundary conditions are needed in order to allow
the
$\varphi=\pi/\beta$ vacuum. The zero mode is present, but the dynamical mass
generation
provides the mass gap which guarantees infrared stability. From the point of
view of analytical
study in the infinite volume limit, this dynamical infrared regulator is the
main source of
difficulties for the perturbative expansion.
With Wilson fermions, a perturbative mass term must be added to compensate the
artificial chiral
symmetry breaking: this device is very expensive in terms of compational
cost~\cite{Curci,Beccaria:GN}
and is not required here.
\section{$1/N$ expansion}
\label{sec:exp}
We discuss the $1/N$ expansion of the model with particular attention to the
quantity
$\langle\sigma\rangle$.
The effective potential for the $\sigma$ vacuum expectation value is
\ba
\label{effg}
\Gamma(\sigma) &=& \frac{\sigma^2}{2\lambda} - \frac{1}{L^2}\log\langle
e^{-S_{\rm int}}\rangle + \\
& & + \frac{1}{2N}
\int \frac{d^2p}{(2\pi)^2} \log\left(\frac{1}{\lambda} + \Pi(p, \sigma)\right)
+
O\left(\frac{1}{N^2}\right).
\nonumber
\ea
Let us consider the various terms appearing in Eq.(\ref{effg}).
The leading order is obtained immediately. For instance, in the bosonic case,
the contribution of $\varphi^\alpha$ factorizes an $N$ factor;
indeed, on a formal level
\ba
\lefteqn{\frac{1}{\cal N}\int {\cal D}\sigma{\cal D}\varphi\exp\left(
-\int d^2x (\frac{N\sigma^2}{2\lambda} +
\frac{1}{2}\left(\partial\varphi^\alpha\right)^2 +
\sum_\alpha \sigma\cos\beta\varphi^\alpha)\right) = } && \ \ \ \nonumber \\
&=& \frac{1}{\cal N}\int {\cal D}\sigma \exp -N\left[
\int d^2x \frac{\sigma^2}{2\lambda} - \ln H(\sigma)
\right]
\ea
where
\be
H(\sigma) = \int {\cal D}\varphi \exp\left(-\int d^2 x
(\frac{1}{2}(\partial\varphi)^2 +
\sigma\cos\beta\varphi) \right) = \frac{1}{{\cal N}^\prime}
\langle \exp\left(-\int d^2x \sigma\cos\beta\varphi\right)\rangle_{\rm free}.
\ee
To the leading order, it is enough to consider a constant $\sigma$ and $\langle
e^{-S_{int}}
\rangle$ is computed with the action
\be
S = \sum_n\frac{N}{2\lambda}\sigma_n^2 + \frac{1}{2}\ \Delta_\mu^+\varphi_n\
\Delta_\mu^+\varphi_n
-\omega\sum_n \sigma\cos\beta\varphi_n
\ee
with only one field $\varphi$ and constant external $\sigma$.

Even the leading order cannot be computed analytically and needs a small Monte
Carlo
simulation. This is unavoidable and is related to the non-triviality of the
bosonized model, also at the tree level; the numerical simulation is much more
efficient than
the usual ones, but analytical predictions become really involved.

The second term in
Eq.(\ref{effg}) comes from integrating a $\varphi$ loop with the propagator
keeping into
account the dynamic mass term induced by the $\sigma$ vacuum expectation value.
Therefore,
$\Pi(p, \sigma)$ is the one particle irreducible $\sigma$ self energy computed
with the
above $N=1$ action in which the dynamical field $\sigma$ is replaced
the solution of the $N=\infty$ saddle point equation $\Gamma^\prime(\sigma) =
0$.
In the same spirit, we can write the leading expression for
$\langle\sigma^2\rangle_c$ which
reads
\be
\langle \sigma^2\rangle_c = \langle \sigma^2\rangle - \langle \sigma\rangle^2 =
\frac{1}{N} \int\frac{d^2p}{(2\pi)^2}\ \frac{1}{1/\lambda + \Pi(p, \sigma)} +
O(1/N^2).
\ee
In~\cite{Beccaria:GNB}, we showed that $\langle\sigma\rangle$ scales correctly
at $N=\infty$.
The next to leading correction is obtained by expanding
both $\Gamma$ and $\langle\sigma\rangle$ in powers of $1/N$. With the notation
\ba
\Gamma(\sigma) &=& \frac{\sigma^2}{2\lambda} +
\Gamma_1(\sigma) + \frac{1}{N} \Gamma_2(\sigma) + O(\frac{1}{N^2}), \\
\sigma &=& \sigma_0 + \frac{1}{N} \sigma_1  + O(\frac{1}{N^2})
\ea
we obtain
\ba
\sigma_0 &=& -\lambda \Gamma_1^\prime(\sigma_0), \\
\sigma_1 &=& -\frac{\Gamma_2^\prime(\sigma_0)}{1/\lambda +
\Gamma_1^{\prime\prime}(\sigma_0)}.
\ea
\section{Scaling and mass spectrum}
\label{sec:RG}
The universal $\beta$ function of the model is defined by
\be
\mu\frac{d\lambda}{d\mu} = -\beta_0\lambda^2 - \beta_1\lambda^3 + O(\lambda^4)
\ee
with
\be
\beta_0 = \frac{N-1}{\pi N}\qquad\qquad \beta_1 = -\frac{N-1}{2\pi^2 N^2}
\ee
Therefore, the two dimensional Gross Neveu model is asymptotically free and
\be
a\sim \left(\lambda\right)^{-\beta_1/\beta_0^2}
\exp\left(-\frac{1}{\beta_0\lambda}\right) =
\left(\frac{N-1}{N}\lambda\right)^\Delta
\exp\left(-\frac{N}{N-1}\frac{\pi}{\lambda}\right)\qquad \Delta =
\frac{1}{2(N-1)}
\ee
where we have written only terms up to $O(1/N)$.
A Renormalization Group analysis of $\Gamma$~\cite{Rossi3}
reveals that $\langle\sigma\rangle$ has an anomalous
scaling and that its bare lattice value behaves as
\be
\label{sigmaasy}
\langle\sigma\rangle \sim \exp\left(-\frac{N}{N-1}\frac{\pi}{\lambda}\right)
\qquad {\rm as}\ \lambda\to 0.
\ee
Both $\langle\sigma^2\rangle_c$ and the action density are afflicted by
perturbative tails and
the comparison of theory with numerical data needs perturbative expansions for
these quantities
which we have not obtained yet.
Apart from the chiral condensate, the mass spectrum is another source of
informations
in order to claim that the above lattice model {\it has} the Gross-Neveu model
as its critical point.
The semiclassical analysis and quantization shows that the spectrum
is composed of particles which are bound states of kinks. The kink state is
very similar to the quantum soliton of the Sine-Gordon
system~(see~\cite{Faddeev} for a
comprehensive review of the quantum treatment of solitons). The
other states fall in multiplets with the mass
formula~\cite{Karowski}
\be
\label{mass1}
\frac{M_n}{M_1} = \frac{\sin n \pi \Delta}{\sin \pi\Delta}
\qquad n = 1,\cdots, N-2
\ee
and the Bethe-Ansatz prediction~\cite{Forgacs}
\be
\label{mass2}
\frac{M_1}{\Lambda_{\overline{\rm MS}}} = \frac{(4e)^\Delta}{\Gamma(1-\Delta)}.
\ee
The bosonization formalism does not encounter particular problems in studying
the
bosonic sector corresponding to even $n$. However, the fermionic states with
odd $n$
cannot be easily determined since they involve non local operators.
The simplest quantity whose scaling can be studied is
\be
r(\lambda) = \frac{M_2(\lambda)}{\Sigma(\lambda)}
\ee
where $M_2(\lambda)$ is the mass extracted from the $\sigma$ propagator and
$\Sigma$ is the
combination
\be
\Sigma = \langle \sigma\rangle
\left(\frac{N-1}{N}\frac{\lambda}{2\pi}\right)^\Delta.
\ee
We remark that the Schwinger-Dyson equations can be used to write
\be
\langle \sigma(x)\sigma(y)\rangle = (\lambda\omega)^2 \langle\frac{1}{N}
\sum_i \cos\beta\varphi^i(x)\ \frac{1}{N}\sum_j \cos\beta\varphi^j(y)\rangle
\ee
and the r.h.s. fluctuates less.
Unfortunately, a measure of $M_4$ would have been too expensive, thus we cannot
check
Eq.(\ref{mass1}). An analytical computation of the $\Lambda$ parameters ratio
$\Lambda/\Lambda_{MS}$
together with Eq.(\ref{mass2}) would permit such a check: we leave this subject
for future work.
The present measure of $M_2$ is a different scaling test and provides
confidence in the correct
mass spectrum.
\section{Numerical results}
\label{sec:num}
We simulated the model using the Hybrid Monte Carlo algorithm~\cite{Duane}.
In Tab.~\ref{results:letter}, we report for completeness our old
results~\cite{Beccaria:GNB} obtained
on a $40^2$ lattice with $N=20$ flavours. We show, at three values of
$\lambda$, the
leading $1/N$ prediction
for $\langle\sigma\rangle$, its $1/N$ correction, the $1/N$ corrected value
considering $N=20$ flavors
and the explicit numerical data including the measure of
$\langle\sigma^2\rangle_c$.
In Tab.~\ref{quantumforcen} we display $F_{\rm qu} = -\Gamma_1^\prime(\sigma)$
over a wide range of
values. The resulting plot is shown in Fig.(\ref{fig:quantumforcen}).
Tab.\ref{tabcorrections:results} shows the $1/N$ expansion of
$\langle\sigma\rangle$ at $N=6$ and with three couples of values of $\lambda$,
$L$
used in Tab.\ref{sigma6}.
The simulation
at $N<\infty$ has been performed at $N=6$ with the results of
Tab.~\ref{sigma6}.
We varied the lattice size until the results were stable; in order to have
uniform finite
size scaling corrections we tried to
have comparable values of the Fisher variable $L \sigma$.
In Fig.(\ref{fig:scaling}) we have compared the numerical data with the
asymptotic prediction
of Eq.(\ref{sigmaasy}) at $N=6$.
Results on the lightest bosonic state can be read in Tab.~\ref{mass6} and in
Fig.(\ref{fig:mass})
we have shown the bare values and the ratio $M_2/\Lambda$ which is compatible
with a constant
value. Within the statistical error, there is no difference in considering the
ratio $r(\lambda)$.
\section{Conclusion}
\label{sec:conc}
We can draw the following conclusions: the direct simulation
of the bosonized model is certainly possible. At comparable values of the
physical quantities,
it is extremely fast with respect to the fermionic realization. This is because
no matrix
inversions are needed. Of course, the conservation of chiral symmetry played an
important role also. The $1/N$ expansion is non trivial and
requires (simple) simulations by itself because the action is not polynomial;
at least for
large $N$ it gives a good agreement. Using the Gross-Neveu model, we have
observed good
scaling properties at relatively low $N$. Of course, all the advantages are
lost if
one wants to study the fermionic sector. The extension of this technique to
other models
is a continuum problem related to the possibility of their bosonization. The
duality approach
seems very interesting in this respect.
\acknowledgments

I wish to thank Prof. Giuseppe Curci for constant encouragement.
I am grateful to
Prof. Paolo Rossi for many stimulating discussions and for a
careful reading of the
manuscript.

\appendix
\section{The Symanzik improvement}
\label{app:symanzik}
Motivated by general considerations prompted in~\cite{Beccaria:GNB}, we
explored the
possibility of simulating the Symanzik improved action~\cite{Symanzik} because
the non
trivial connectivity of the action is computationally irrelevant.
We can follow the general idea of reducing, at least at the classical level,
the errors
introdued by the lattice discretization.
We replace the kinetic term
\be
-\frac{1}{2} \varphi_n (\varphi_{n+\mu} - 2 \varphi_n + \varphi_{n-\mu})
\ee
with the combination
\be
\frac{1}{24}\varphi_n\left(\varphi_{n+2\mu}-16\varphi_{n+\mu}
+30\varphi_n-16 \varphi_{n-\mu}+\varphi_{n-2\mu}
\right)
\ee
The bosonic kinetic action is therefore
\be
S = \frac{1}{24}\sum_{n\mu}\left[16(\varphi_{n+\mu}-\varphi_n)^2
-(\varphi_{n+2\mu}-\varphi_n)^2
\right].
\ee
The interaction term depends on the infrared behaviour of the tadpole which
must be
computed with the improved propagator, otherwise the correct slope in the
asymptotic
scaling law is not obtained.
The small mass expansion of the tadpole is
\be
\int\frac{d^2p}{(2\pi)^2}\frac{1}{\hat{p}^2 + \hat{p}^4/12 + \alpha^2} =
-\frac{1}{4\pi}\ln\mu^2 + C_S \qquad
\hat{p}^4\equiv \sum_\mu \hat{p}_\mu^4
\ee
where we have defined
\be
C_S = \frac{5}{4\pi}\ln 2
-\frac{1}{12}\int\frac{d^2p}{(2\pi)^2}\frac{\hat{p}^4}
{\hat{p}^2(\hat{p}^2+ \hat{p}^4/12)} = \frac{5}{4\pi}\ln 2 + I_S.
\ee
Therefore the constant $\omega$ is altered by a factor
\be
\omega = \frac{e^\gamma 2^{3/2}}{\pi}\exp(2\pi I_S).
\ee
A numerical evaluation of $I_S$ gives (I show only the first digits)
\be
I_S \simeq -0.04717 \rightarrow \omega_S \simeq 1.19223
\ee
The improvement lowers $\omega$ and presumably also the $\Lambda$ parameter
worsening the scaling
behaviour.
This conclusion is confirmed by preliminary numerical data at
small $N$. Therefore we did not find any obvious advantage in using the tree
level improved action.
\begin{table}
\caption{Summary of the $N=20$ results.}
\vskip 0.5truecm
\label{results:letter}
\begin{tabular}{c|cccccc}
$\lambda$ & $\sigma_0$ & $\sigma_1$ & $\sigma_0 + \sigma_1/20$ & $\langle
\sigma\rangle$
& $\langle \sigma^2\rangle_c$ & $R$ \\
\tableline
0.49000 & 0.1488 & $-$           & $-$            & $0.090(2)$ & $0.0606(3)$ &
$77(2)$\\
0.51905 & 0.2000 & $-1.7(2)$ & $0.12(1)$ & $0.127(1)$ & $0.0681(6)$ &
$74.6(6)$\\
0.54376 & 0.2500 & $-2.1(2)$ & $0.15(1)$ & $0.166(1)$ & $0.0744(6)$ &
$72.7(4)$\\
0.56876 & 0.3000 & $-2.4(2)$ & $0.18(1)$ & $0.209(1)$ & $0.0807(7)$ &
$70.0(3)$\\
\end{tabular}
\end{table}
\begin{table}
\caption{${\rm F}_{\rm qu}(\sigma)$ at $L=40$.}
\vskip 0.5truecm
\label{quantumforcen}
\begin{tabular}{ccc|cc}
$\sigma$ & $\omega\langle\cos\beta\varphi\rangle$ & &
$\sigma$ & $\omega\langle\cos\beta\varphi\rangle$\\
\tableline
0.002   & 0.004(2) & & 0.20    & 0.3853(4) \\
0.005   & 0.012(2) & & 0.25    & 0.4598(3) \\
0.01    & 0.026(2) & & 0.30    & 0.5275(3) \\
0.02    & 0.050(2) & & 0.40    & 0.6502(3) \\
0.03    & 0.075(2) & & 0.50    & 0.7560(3) \\
0.04    & 0.095(1) & & 0.60    & 0.8464(2) \\
0.05    & 0.119(1) & & 0.70    & 0.9243(2) \\
0.06    & 0.140(1) & & 1.00    & 1.0988(2) \\
0.10    & 0.2182(6)& & 2.00    & 1.3515(2) \\
\end{tabular}\end{table}

%

\begin{table}
\caption{Results of the $1/N$ expansion.}
\vskip 0.5truecm
\label{tabcorrections:results}
\begin{tabular}{ccccc}
L & $\lambda$ & $\langle\sigma\rangle_0$ &
$\langle\sigma\rangle_1$ & $\langle\sigma^2\rangle_{c, 0}$ \\
\tableline
40  & 0.800   & 0.7854(4) & $0.4784 \pm 0.14$ & 0.2543(2)\\
50  & 0.700   & 0.5809(4) & $0.3433 \pm 0.11$ & 0.2573(2)\\
60  & 0.675   & 0.5280(4) & $0.2301 \pm 0.11$ & 0.2577(2)\\
\end{tabular}\end{table}

\begin{table}
\caption{Numerical data at $N=6$.}
\vskip 0.5truecm
\label{sigma6}
\begin{tabular}{cccccc}
L & $\lambda$ & $\langle\sigma\rangle$ & $\langle\sigma^2\rangle_c$ &
$\langle S \rangle$/N & $L\cdot \langle\sigma\rangle$\\
\tableline
40  & 0.800   & 0.369(2)   & 0.510(2)    & 0.1770(8) & 15  \\
40  & 0.700   & 0.200(1)   & 0.3930(7)   & 0.3137(4) & 8   \\
50  & 0.700   & 0.200(1)   & 0.3930(6)   & 0.3135(4) & 10  \\
60  & 0.675   & 0.166(1)   & 0.3640(6)   & 0.3412(5) & 10  \\
60  & 0.650   & 0.132(2)   & 0.3360(8)   & 0.3669(6) & 7.9 \\
76  & 0.650   & 0.136(2)   & 0.3351(7)   & 0.3658(6) & 10  \\
80  & 0.625   & 0.108(1)   & 0.3077(3)   & 0.3888(3) & 8.6 \\
80  & 0.600   & 0.090(1)   & 0.2808(3)   & 0.4077(3) & 7.2 \\
110 & 0.600   & 0.0845(7)  & 0.2814(2)   & 0.4088(2) & 9.3 \\
120 & 0.575   & 0.0635(1)  & 0.2567(1)   & 0.4267(1) & 7.6 \\
\end{tabular}\end{table}

\begin{table}
\caption{Mass measurement at $N=6$.}
\vskip 0.5truecm
\label{mass6}
\begin{tabular}{ccc}
L & $\lambda$ & $M_2$ \\
\tableline
60 & 0.675  & 0.25(4) \\
76 & 0.650  & 0.20(4) \\
80 & 0.625  & 0.16(3) \\
\end{tabular}\end{table}
%
%
\begin{figure}[htbp]
\begin{center}
\mbox{\psfig{file=veff.ps,width=3.0truein,angle=-90}}
\end{center}
\caption{$F_{\rm qu}(\sigma)$ on the $40^2$ lattice.}
\label{fig:quantumforcen}
\end{figure}
\newpage
%
%
\begin{figure}[htbp]\begin{center}
\mbox{\psfig{file=scalingnew.ps,width=3.0truein,angle=-90}}
\end{center}
\caption{$O(1/N)$ asymptotic scaling.}
\label{fig:scaling}
\end{figure}
\newpage
\begin{figure}[htbp]\begin{center}
\mbox{\psfig{file=mass.ps,width=3.0truein,angle=-90}}
\end{center}
\caption{Scaling of $M_2$.}
\label{fig:mass}
\end{figure}
\newpage
\end{document}